\documentclass[sigconf]{acmart}

\usepackage{framed}
\usepackage{graphicx}
\usepackage{acmart-taps}
\usepackage{array}
\usepackage{boldline} % 用于表头粗线
\usepackage{tabularx} % 用于指定表格的宽度
\usepackage{algorithm}
\usepackage{algpseudocode}
\usepackage{afterpage}
\usepackage{multirow}
\usepackage{xcolor} % 加载xcolor宏包
\usepackage{colortbl} % 加载colortbl宏包
\usepackage{bm} 
\definecolor{shadecolor}{gray}{0.9}

\definecolor{lb}{RGB}{200,200,200} 
\usepackage{amsthm}
\usepackage{enumitem}
\usepackage{subcaption} % 引入宏包
\usepackage{caption}
\usepackage{microtype}
\copyrightyear{2024}
\acmYear{2024}
\setcopyright{acmlicensed}
\acmConference[Internetware 2024]{15th Asia-Pacific Symposium on Internetware}{July 24--26, 2024}{Macau, Macao}
\acmBooktitle{15th Asia-Pacific Symposium on Internetware (Internetware 2024), July 24--26, 2024, Macau, Macao}
\acmDOI{10.1145/3671016.3671388}
\acmISBN{979-8-4007-0705-6/24/07}

\begin{document}
\title{DFEPT: Data Flow Embedding for Enhancing Pre-Trained Model Based Vulnerability Detection}

\author{Zhonghao Jiang}
\email{zhonghaojiang@cqu.edu.cn}
\affiliation{%
\institution{School of Big Data and Software Engineering, Chongqing University}
\city{}
\country{China}}

\author{Weifeng Sun}
\email{weifeng.sun@cqu.edu.cn}
\affiliation{%
\institution{School of Big Data and Software Engineering, Chongqing University}
\city{}
\country{China}}

\author{Xiaoyan Gu}
\email{xiaoyan.gu@stu.cqu.edu.cn}
\affiliation{%
\institution{School of Big Data and Software Engineering, Chongqing University}
\city{}
\country{China}}

\author{Jiaxin Wu}
\email{wujiaxin@stu.cqu.edu.cn}
\affiliation{%
\institution{School of Big Data and Software Engineering, Chongqing University}
\city{}
\country{China}}

\author{Tao Wen}
\email{wente@cqu.edu.cn}
\affiliation{%
\institution{School of Big Data and Software Engineering, Chongqing University}
\city{}
\country{China}}

\author{Haibo Hu}
\email{haibo.hu@cqu.edu.cn}
\affiliation{%
\institution{School of Big Data and Software Engineering, Chongqing University}
\city{}
\country{China}}

\author{Meng Yan}
\email{mengy@cqu.edu.cn}
\authornote{Meng Yan is the corresponding author.}
\affiliation{%
\institution{School of Big Data and Software Engineering, Chongqing University}
\city{}
\country{China}}

\begin{abstract}
		Software vulnerabilities represent one of the most pressing threats to computing systems.
		Identifying vulnerabilities in source code is crucial for protecting user privacy and reducing economic losses. 
		Traditional static analysis tools rely on experts with knowledge in security to manually build rules for operation, a process that requires substantial time and manpower costs and also faces challenges in adapting to new vulnerabilities. 
		The emergence of pre-trained code language models has provided a new solution for automated vulnerability detection. 
		However, code pre-training models are typically based on token-level large-scale pre-training, which hampers their ability to effectively capture the structural and dependency relationships among code segments. 
		In the context of software vulnerabilities, certain types of vulnerabilities are related to the dependency relationships within the code. 
		Consequently, identifying and analyzing these vulnerability samples presents a significant challenge for pre-trained models.
		
		In this paper, we propose a data flow embedding technique to enhance the performance of pre-trained models in vulnerability detection tasks, named DFEPT, which provides effective vulnerability data flow information to pre-trained models. 
		Specifically, we parse data flow graphs (DFG) from function-level source code, and use the data type of the variable as the node characteristics of the DFG. 
		By applying graph learning techniques, we embed the data flow graph  and incorporate relative positional information into the graph embedding using sine positional encoding to ensure the completeness of vulnerability data flow information. 
		Our research shows that DFEPT can provide effective vulnerability semantic information to pre-trained models, achieving an accuracy of 64.97\% on the Devign dataset and an F1-Score of 47.9\% on the Reveal dataset.
		Compared with the pre-trained model that is only fine-tuned, the performance increases by 1.96\%-17.26\%.
		
	\end{abstract}

\renewcommand{\shortauthors}{Zhonghao Jiang et al.}

\sloppy
	\maketitle	
\vspace{-0.2cm}
\section{Introduction}
In recent times, software vulnerabilities have emerged as a pivotal concern within the realm of software security \cite{neuhaus2007predicting,zhou2019devign}. 
Due to security vulnerabilities, the personal information of many internet users has been compromised, with common reports exposing incidents involving millions of data breaches \cite{wiki_data_breaches_2021}.
Furthermore, the Common Vulnerabilities and Exposures (CVE) reports from 2016 to 2021 \cite{cve_vulnerabilities_2021} indicate that the number and variety of vulnerabilities are growing at an alarmingly rapid rate.
These cybersecurity attacks have brought incalculable harm and losses to the economy and society \cite{fu2022linevul,thapa2022transformer}. 
%Therefore, there is an urgent need to develop efficient vulnerability detection tools to identify potential vulnerabilities in source code.
Consequently, it's imperative to create effective tools for detecting vulnerabilities, aiming to pinpoint potential security flaws in the source code.

Early methods \cite{neuhaus2007predicting,nguyen2010predicting} of vulnerability detection rely on the prior knowledge of domain experts \cite{shin2010evaluating}. 
Experts have to manually review large amount of source code and set specific rules or manually extract features for detection. 
These approaches are costly and inefficient because manually setting rules and extracting features requires a significant amount of time and effort. 
Moreover, these rules need constant maintenance and updating to adapt to all the vulnerabilities in a codebase, which is not feasible with the traditional approach.

The rapid development of artificial intelligence has driven software companies to invest in vulnerability detection tools based on deep learning \cite{lu2021codexglue, zheng2021d2a}. 
These tools have achieved commendable performance, even surpassing traditional static analysis methods \cite{cesare2013detecting,ding2022velvet,li2018vuldeepecker}. 
Currently, deep learning-based vulnerability detection tools mainly fall into two categories:
(1) Graph-based Models: 
%These methods aim to leverage static analysis to extract various dependency relationships in source code, such as Abstract Syntax Trees (AST), Control Flow Graphs (CFG), Data Flow Graphs (DFG), and Program Dependence Graphs (PDG). 
These approaches seek to utilize static analysis to uncover different types of dependency relationships within source code, including Abstract Syntax Trees (AST), Control Flow Graphs (CFG), Data Flow Graphs (DFG), and Program Dependence Graphs (PDG).
%They use graph learning to aggregate structural and semantic information from the source code for detection.
These techniques employ graph learning to integrate both structural and semantic insights derived from the source code for detection.
For example, Devign \cite{zhou2019devign} uses a Programming Language (PL) parser to extract multifaceted graph information for classification. 
DeepWukong \cite{cheng2021deepwukong} first generates Program Dependence Graphs by considering control flow and data dependencies, then extracts subgraphs, named XFGs,  from system API calls or operators, and finally inputs XFG to train slice-level vulnerability detection models.
(2) Sequence-based Models: 
These methods treat source code as sequences akin to natural language, employing prevalent natural language processing techniques to identify vulnerabilities \cite{lu2021codexglue}. 
The emergence of pre-trained and fine-tuning techniques has become a popular learning paradigm \cite{devlin2018bert}. 
Inspired by models like BERT \cite{devlin2018bert}, researchers sequentially propose pre-trained models such as CodeBERT \cite{feng2020codebert}, GraphCodeBERT \cite{guo2020graphcodebert}, UniXcoder \cite{guo2022unixcoder}, and LineVul \cite{fu2022linevul}. 
These models have successively demonstrated good performance in vulnerability detection tasks \cite{steenhoek2023empirical}.

However, in practical applications, the aforementioned methods usually exhibit several limitations:
(1) During the embedding process, the node features of the graph are typically slice-level code segments. 
Unfortunately, text or token information unrelated to vulnerabilities, such as variable names and function names, is also embedded, introducing additional noise and interference \cite{steenhoek2023dataflow}.
(2) The relative position of nodes in the code structure graph is critical to the emergence of vulnerabilities.
The same code segment may be marked with different security statuses in different contexts \cite{nguyen2022regvd}. 
However, graph learning overlooks the relative position of nodes.
(3) When detecting vulnerabilities, it is challenging for these models to understand the semantics of vulnerabilities in the code \cite{steenhoek2023dataflow, wang2023defecthunter}.
%Although pre-trained models can be fine-tuned for specific downstream tasks, learning only at the text level makes it difficult to capture the inherent structure of the code, such as data flow and control flow.
While pre-trained models can be adapted for particular tasks, focusing solely on textual analysis hampers their ability to grasp the code's intrinsic structure, like data flow and control flow.

To overcome the aforementioned limitations, we propose DFEPT, a data flow embedding technique designed to enhance the performance of pre-trained models in vulnerability detection tasks. 
DFEPT provides pre-trained models with the semantic information of data flows in code structures and the relative position information of nodes, thereby achieving better performance. 
Specifically, DFEPT utilizes open-source code parsing tools to extract ASTs, and then parses data flow graphs from the ASTs. 
To accurately identify each specific data flow, we use only the data types of variables to initialize node features, avoiding the introduction of extraneous information. 
To incorporate the relative position features of nodes, we adopt the sinusoidal positional encoding from the Transformer \cite{vaswani2017attention} architecture to encode graph embeddings, providing the classifier with a larger receptive field.
We concatenate the extracted data flow information with the output of the pre-trained model and input them into the classifier for detection. 
This not only ensures the strong performance of pre-trained models in vulnerability detection but also supplements the models with effective vulnerability semantics, enhancing their performance and effectively addressing the three issues mentioned above.

We evaluate the performance of DFEPT combined with four pre-trained models: CodeBERT \cite{feng2020codebert}, GraphCodeBERT \cite{guo2020graphcodebert}, CodeT5 \cite{wang2021codet5}, and UniXcoder \cite{guo2022unixcoder}, on two popular datasets, Devign and Reveal, and compare it with seven state-of-the-art methods. 
The experimental results show that the performance of all pre-trained models are significantly improved when combined with DFEPT, demonstrating its generalizability. 
%Furthermore, by combining CodeT5 with DFEPT, we achieve a leading performance with an accuracy of 64.53\% and an F1-Score of 61.22\% on the Devign dataset, and an accuracy of 92.92\% and an F1-Score of 47.9\% on the Reveal dataset, surpassing all baseline methods. 
Furthermore, integrating CodeT5 with DFEPT results in superior performance, achieving an accuracy of 64.53\% and an F1-Score of 61.22\% on the Devign dataset, along with an accuracy of 92.92\% and an F1-Score of 47.9\% on the Reveal dataset, outperforming all comparative baseline approaches.
We also study the impact of different data flow graph embedding and pooling methods on DFEPT's performance.
The results show that DFEPT performs the best when using Graph Convolutional Neural Networks and united pooling (a combination of sum pooling and max pooling). 
Even with the least effective embedding methods, DFEPT still provides valuable information to pre-trained models to enhance vulnerability detection performance.

In summary, the main contributions of this paper are as follows:
\begin{itemize}[topsep=0pt,noitemsep, left=0pt]
\item \textbf{Approach.} We propose a data flow embedding technique, which uses the data types of variables to identify data flows and applies graph learning and sine positional encoding to generate effective semantic information and relative positional information about vulnerabilities. 
This enables pre-trained models to capture semantics and data flow patterns in vulnerabilities.
	
\item \textbf{Study.} 
%We evaluate the performance of DFEPT on two datasets and with four pre-trained models, comparing it against seven baseline methods. 
%	We also conduct ablation experiments to verify the necessity of each component of DFEPT. 
%	The results demonstrate that the information provided by DFEPT effectively enhances the ability of pre-trained models to detect vulnerabilities. 
We assess DFEPT's effectiveness using two datasets and four pre-trained models, benchmarking it against seven baseline strategies. 
Additionally, we perform ablation studies to confirm the essential role of each DFEPT component. 
The findings reveal that, compared to merely fine-tuned pre-trained models, DFEPT's insights enhance the models' ability to identify vulnerabilities, with a performance improvement ranging from 1.96\% to 17.23\%.
Each component of DFEPT contributes valuable information.
	
\item \textbf{Open Science.} We combine DFEPT with common pre-trained models to create high-performing and efficient vulnerability detection models. 
By combining with CodeT5 \cite{wang2021codet5}, we surpass the most advanced vulnerability detection performance. 
The source code of DFEPT is publicly available\footnote{\url{https://github.com/GCVulnerability/DFEPT}}.
\end{itemize}

\begin{figure*}[th!]
	\centering
	\includegraphics[width=1\textwidth]{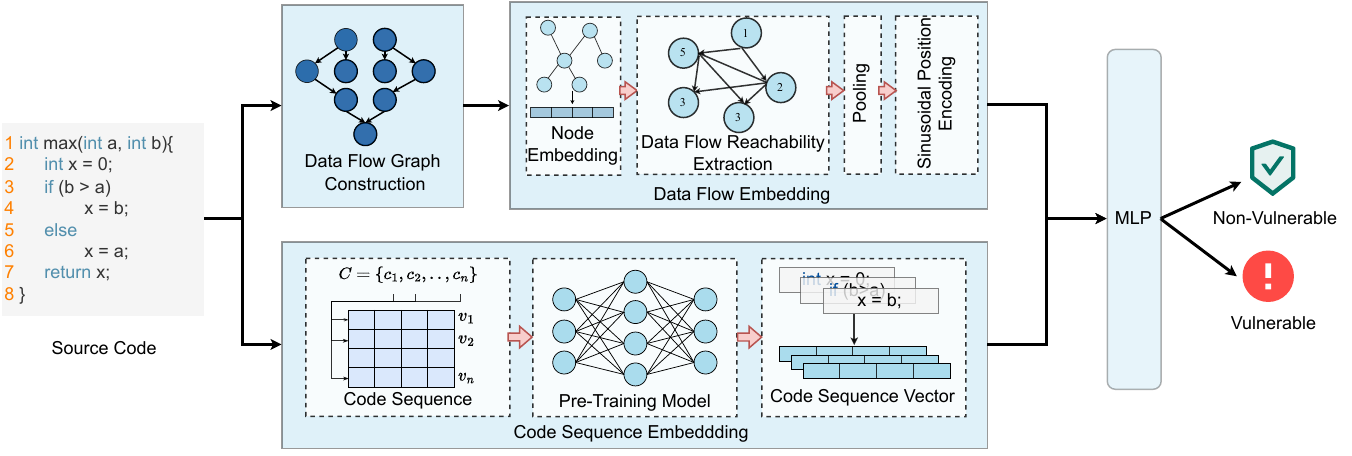}
	\caption{Overview of DFEPT.}
	\label{Fig1}
\end{figure*}	
\vspace{-0.2cm}
\section{Preliminaries}
\subsection{Data Flow Graph}
%A data flow graph is a code graph structure that represents the dependency relationships between variables. 
A data flow graph represents a structural code graph that depicts the dependencies and interactions among variables.
Its nodes represent variables, and the edges indicate the sources of these variables. 
For source code that implements the same functionality, different ASTs might be parsed due to variations in programmer preferences and naming conventions.
However, these pieces of code will have the same data flows. 
Moreover, data flow can identify long-distance dependencies arising from the use of the same variables or functions in lengthy codes. 

Data flow graphs are not only helpful in understanding the structure and data dependencies of a program, but they are also crucial for vulnerability detection. 
By mapping the flow and interaction of variables within a program, data flow graphs can reveal potential security vulnerabilities such as buffer overflows, injection attacks, or unauthorized data access. 
Security analysts can use these diagrams to trace the flow of untrusted input data, thereby identifying weaknesses that could be exploited by attackers to compromise program security.
Therefore, data flow graphs have become an indispensable tool in software security analysis, making the detection of vulnerabilities more intuitive and efficient.

\vspace{-0.3cm}
\subsection{Graph Reachability  and Graph Learning }
In graph theory, graph reachability is one of the most important characteristics of a graph, which measures whether there is a path between any two nodes. 
The commonly used representation method of reachability is the reachability matrix, which can be regarded as starting from an initial node and gradually aggregating reachability information to neighboring nodes. 
When calculating the reachability matrix, we incorporate information aggregated from neighbor nodes each time to update the reachability matrix state. 
Graph learning has similar characteristics. 
Graph learning starts from the representation of initial nodes, executes a specific message propagation algorithm \cite{gilmer2017neural} and propagates node information through a fixed number of iterations. 
In each iteration, each node aggregates the feature vectors of neighboring nodes and then updates the hidden state. Their similar state equations are as follows:
$$\text{reachability matrix:}\ P^{(0)}=\mathcal{A}^0 \quad P^{(n)}=P^{(n-1)} \vee \mathcal{A}^{(n-1)}$$
$$\text{graph learning:}\ H^{(0)}=\mathcal{X} \quad H^{(n)}= GNN(H^{(n-1)}, \mathcal{A})$$
where $\mathcal{A}$ is the reachability matrix of the graph. $\mathcal{X}$ is the node feature embedding of the graph. $P$ and $H$ represent the intermediate state of the reachability matrix and the hidden state in graph learning respectively.

Some software vulnerabilities are closely related to the reachability of data flow graphs, such as controlling pointer dereferences and using uninitialized variables \cite{cesare2013detecting}. 
These vulnerabilities are partly caused by the incorrect data flow reaching the target variable node or the correct data flow being unreachable. 
Therefore, using graph learning to perform reachability analysis on data flow graphs is helpful for vulnerability detection.

\section{Methodology}
%In this section, we will introduce a new data flow embedding method for enhancing pre-trained model based vulnerability detection, called DFEPT, to identify whether a function-level code is vulnerable or not.
In this section, we will introduce DFEPT, a novel data flow embedding technique designed to enhance vulnerability detection in pre-trained models, aimed at determining the vulnerability status of function-level code.
The complete workflow of DFEPT is shown in Figure \ref{Fig1}.
DFEPT initially constructs data flow graphs from function-level source code as described in Section \ref{3.1}. 
Then, the constructed data flow graph nodes and structures are embedded as detailed in Section \ref{3.2}. 
Subsequently, the embedded graph feature vectors are applied with the sinusoidal positional encoding introduced in Section \ref{3.3}. 
Finally, after concatenation with the code sequence embedding presented in Section \ref{3.4}, they are fed into an Multilayer Perceptron (MLP) for classification and detection.

\vspace{-0.3cm}
\subsection{Data Flow Graph Construction}\label{3.1}

\begin{figure}[h!]
	\centering
	\resizebox{0.45\textwidth}{!}{\includegraphics[width=1\textwidth]{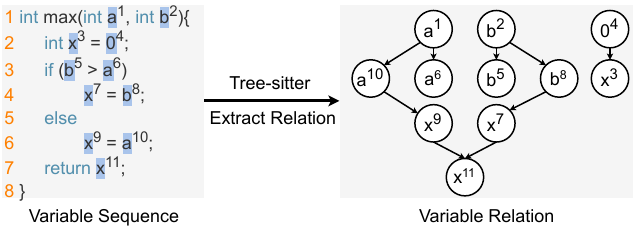}}
	\caption{The process of building a data flow graph.}
	\label{Fig2}
\end{figure}

As is shown in Figure \ref{Fig2}, for the source code given in Figure \ref{Fig1}, represented as $C = \{c_1, c_2,...,c_n\}$, we first utilize a standard source code parsing tool, tree-sitter \footnote{\url{https://github.com/tree-sitter/tree-sitter}}, to construct the AST of $C$. 
Subsequently, we identify a sequence of variables $\mathcal{V}=\{v_1, v_2, ... v_n\}$ from the leaf nodes of the AST and use them as nodes for the DFG. 
We then extract relationships between nodes from the AST and construct directed edges $e=\left \langle  v_i, v_j\right \rangle $, indicating that the value of the j-th variable originates from the i-th variable. 
We represent the set of directed edges as $\mathcal{E}=\{e_1, e_2, ... e_n\}$. 
The graph $\mathcal{G}(C)=(\mathcal{V},\mathcal{E})$ represents the data flow graph of the source code C.

\vspace{-0.2cm}
\subsection{Data Flow Embedding}\label{3.2}
For a given DFG $\mathcal{G}$, an important question is how to form a code graph embedding that encapsulates the rich semantic information of $\mathcal{G}$. 
We divide the data flow graph embedding process into three parts: node embedding, graph neural network embedding, and pooling.
\subsubsection{Node Embedding}
To accurately and conveniently identify different data flows on the same Data Flow Graph, and to extract code structure information without affecting the judgment of the code sequence, we opt to use the variable type (e.g. int, float etc.) on the node instead of the variable name as the node feature for embedding. 
Specifically, we initialize the embedding layer using the weights of an embedding layer from a pre-trained model that will be used in section \ref{3.4}. 
Then, we tokenize the node features and utilize the embedding layer to obtain the word vector corresponding to the data type of the node.
Assuming that the data type corresponding to the variable on node \( v \) is \( data\_type \), then the node feature embedding \( E(v) \) for node \( v \) is as follows:
\begin{equation}
	E(v) = Embedding\_Layer(Tokenize(data\_type))
\end{equation}
It's important to note that for the same node, using different pre-trained models may yield different embeddings. This variation is due to the fact that different pre-trained models often correspond to different tokenizers and embedding layers.

\subsubsection{Data Flow Reachability Extraction}

%Graph Neural Networks (GNNs) excel in utilizing information propagation mechanisms.
%They recursively aggregate the feature vectors of neighboring nodes to update the node features \cite{kipf2016semi,scarselli2008graph}.
%For a given data flow graph $\mathcal{G}$, let $\mathcal{A}$ represent the adjacency matrix of the data flow graph and $\mathcal{X}$ represent the matrix of node features after node embedding. 
%The basic working equation of the GNN can be represented as follows:
Graph Neural Networks (GNNs) stand out for their effective information propagation mechanisms, where they iteratively consolidate feature vectors from adjacent nodes to refine each node's features \cite{kipf2016semi}. 
Given a data flow graph, denoted as \(\mathcal{G}\), with \(\mathcal{A}\) representing its adjacency matrix and \(\mathcal{X}\) the matrix of node features post-embedding, the fundamental operational formula for a GNN is outlined below:
\begin{equation}
	\mathbf{H}^{(0)} = \mathcal{X}
\end{equation}
\begin{equation}
		\mathbf{H}^{(k+1)} = GNN(\mathcal{A},\mathbf{H}^{(k)})
\end{equation}
where $\mathbf{H}^{(k)}$ is the matrix representation of nodes at the k-th iteration.

Recent studies have proposed various GNNs and their variants. 
We employ Graph Convolutional Neural Networks (GCNs) and Gated Graph Neural Networks (GGNNs) to aggregate node features, thereby generating the embedding vectors for the data flow graph.
Formally, GCN used in our method can be presented as follows:
\begin{equation}
	\mathbf{h}_v^{(k+1)}=ReLU(\sum{a_{v,u}\mathbf{W}^{(k)}\mathbf{h}_u^{(k)} }),\quad\forall v\in \mathcal{V} 
\end{equation}
%where $a_{v,u}$ is an edge constant between nodes $v$ and $u$ in the Laplacian re-normalized adjacency matrix $\mathbf{D}^{-\frac{1}{2}}\mathcal{A}\mathbf{D}^{\frac{1}{2}}$, wherein $\mathbf{D}$ is the diagonal node degree matrix of $\mathcal{A}$, $\mathbf{W}^{(k)}$ is a weight matrix.
In this formula, \(a_{v,u}\) represents the edge constant between nodes \(v\) and \(u\) within the Laplacian re-normalized adjacency matrix \(\mathbf{D}^{-\frac{1}{2}}\mathcal{A}\mathbf{D}^{\frac{1}{2}}\), where \(\mathbf{D}\) is the diagonal matrix denoting the degree of nodes in \(\mathcal{A}\), and \(\mathbf{W}^{(k)}\) is a matrix of weights associated with the edges.

The Gated Graph Neural Network applies the gating concept of GRUs (Gated Recurrent Units) \cite{cho2014learning} to graph networks.
It recursively updates states over a fixed number of time steps. 
The principle of the GGNN we use is as follows:
\begin{equation}
	\mathbf{a}_v^{(k+1)}=\sum{a_{v,u}\mathbf{h}_u^{(k)} }
\end{equation}
\begin{equation}
	\mathbf{z}_v^{(k+1)} = Sigmoid(\mathbf{W}^z\mathbf{a}_v^{(k+1)}+\mathbf{U}^z\mathbf{h}_v^{(k)})
\end{equation}
\begin{equation}
	\mathbf{r}_v^{(k+1)} = Sigmoid(\mathbf{W}^r\mathbf{a}_v^{(k+1)}+\mathbf{U}^r\mathbf{h}_v^{(k)})
\end{equation}
\begin{equation}
\widetilde{\mathbf{h}}_v^{(k+1)}  =ReLU(\mathbf{W}^o\mathbf{a}_v^{(k+1)}+\mathbf{U}^o(\mathbf{r}_v^{(k+1)}\odot \mathbf{h}_v^{(k)} ))
\end{equation}
\begin{equation}
\mathbf{h}_v^{(k+1)} =(1-\mathbf{z}_v^{(k+1)})\odot \mathbf{h}_v^{(k)}+\mathbf{z}_v^{(k+1)}\odot \widetilde{\mathbf{h}}_v^{(k+1)} 
\end{equation}
where $\mathbf{z}$ and $\mathbf{r}$ are the update and reset gates and $\odot$ is the element-wise multiplication.

\subsubsection{Pooling}
%Through GCN or GGNN, we obtain the hidden state representation of each node in the Data Flow Graph. 
By employing GCN or GGNN, we can derive the hidden state representation for each node within the Data Flow Graph.
Before performing classification tasks, it is necessary to aggregate the learned graph features into a single feature vector.
First, we use a soft attention mechanism on the nodes to generate the final feature vector \( \mathbf{e}_v \) for node \( v \). 
Then, we utilize a united pooling method proposed by Nguyen et al. \cite{nguyen2022regvd} to generate the embedding \( \mathbf{e}_g \) for the data flow graph.
The united pooling method combines the advantages of sum pooling and max pooling. 
It uses max pooling to leverage more information about node representations and employs sum pooling to generate more precise graph classification performance \cite{xu2018powerful}.
The generation equations for \( \mathbf{e}_g \) are as follows:
\begin{equation}
\mathbf{e}_v = Sigmoid(\mathbf{W}^T\mathbf{h}_v^{(k)}+\mathbf{b})\odot ReLU(\mathbf{W}\mathbf{h}_v^{(k)}+\mathbf{b})
\end{equation}
\begin{equation}
\mathbf{e}_g = \sum_{v\in \mathcal{V} } \mathbf{e}_v \odot MaxPool\{\mathbf{e}_v\}
\end{equation}

\vspace{-0.2cm}
\subsection{Sinusoidal Position Encoding}\label{3.3}
%DefectHunter 3.C
To better adopt the information of the relative positions of nodes in the data flow graph of lengthy codes, we treat the pooled graph vector as a sequence and then use sinusoidal positional encoding to provide the classifier with fundamental information about the arrangement of elements. 
Sinusoidal positional encoding is first applied in the Transformer \cite{vaswani2017attention} model, where it integrates the position information of the sequence into the original sequence. 
This method of enhancing the model's input by injecting word order has achieved good performance in semantic understanding.
The sinusoidal positional encoding is generated according to the following formula:
\begin{equation}
\begin{cases}
	pos_{2i} = sin(\frac{pos}{10000^{2i/d}} ) \\
	pos_{2i+1} = cos(\frac{pos}{10000^{2i/d}} )
\end{cases}
\end{equation}
where $i$ is the position within the sequence, $d$ is the dimension of the sinusoidal encoding, and $pos_i$ corresponds to the sinusoidal encoding of the i-th token.

By summing sinusoidal positional encoding, we obtain the final embedding vector of the data flow graph, which encompasses code structural information, code semantic information, and relative positional information.
This will provide the pre-trained model with more effective information for its processing.

\vspace{-0.2cm}
\subsection{Code Sequence Embedding}\label{3.4}
%DefectHunter 3.B
The core of Code Sequence Embedding is to treat code segments as natural language sequences and to generate embedding vectors for the entire code block using a pre-trained model.
%First, we tokenize the code segment and input each token into the pre-trained model to obtain word embedding vectors. 
%Then, through different pre-trained model architectures, we transform the word vectors into sentence vectors containing the semantic information of the entire code segment. 
Initially, we tokenize the code segment and feed each token into the pre-trained model to generate word embedding vectors. 
Subsequently, utilizing various pre-trained model architectures, these word embedding vectors are transformed into sentence vectors that encapsulate the semantic essence of the entire code segment.
Finally, we concatenate the generated sentence vectors with the data flow embeddings produced in Section \ref{3.3}, input them into a classifier with two fully connected layers, and construct a vulnerability detection model by minimizing the cross-entropy \cite{zhou2019devign} loss function.

Assuming the pre-trained model we use is referred to as $Model$, the representation of Code Sequence Embedding $\mathbf{e}_c$ for the source code sequence \( C \) is as follows: 
\begin{equation}
	\mathbf{e}_c = Model(C) 
\end{equation}

\vspace{-0.2cm}
\section{Experiment Setup}
%We have conducted extensive experiments to evaluate the performance of DFEPT. 
%Section \ref{4.1} will introduce the data set used in the experiment. 
%Section \ref{4.2} will introduce the baseline method for experimental comparison.
%Section \ref{4.3} will introduce our performance evaluation metrics. 
%Section \ref{4.4} will introduce the research questions we set. 
%DFEPT has been implemented using Python 3.9.18 \cite{python_website} and PyTorch 2.1.0 \cite{pytorch_website}. 
%All experiments are carried out on an Ubuntu 20.04.1 LTS server with 4 Nvidia Geforce RTX 3090 GPUs, a 32-core processor, and 256GB of physical memory.
We carry out comprehensive experiments to evaluate DFEPT's efficacy. 
The dataset utilized in the study will be detailed in Section \ref{4.1}, while Section \ref{4.2} will cover the baseline methods used for comparison. 
Section \ref{4.3} will describe the metrics employed for performance evaluation, and Section \ref{4.4} will outline the research questions we aim to address. 
DFEPT is implemented using Python 3.9.18 \cite{python_website} and PyTorch 2.1.0 \cite{pytorch_website}. 
All testing is conducted on an Ubuntu 20.04.1 LTS server equipped with 4 Nvidia Geforce RTX 3090 GPUs, a 32-core CPU, and 256GB of RAM.
\vspace{-0.2cm}
\subsection{Dataset}\label{4.1}
To verify the effectiveness of DFEPT, we fine-tune the pre-trained model on two widely used vulnerability datasets, including Devign \cite{zhou2019devign} and Reveal \cite{chakraborty2021deep}. 
Each data set has been used in the State of the art research \cite{steenhoek2023empirical, lu2021codexglue, chakraborty2021deep}.
The overview of the dataset and split details are shown in Table \ref{dataset}.

\textbf{Devign.} 
Devign is constructed by Zhou et al. \cite{zhou2019devign}, extracting 27318 manually labeled function fragments as vulnerable or non-vulnerable from two large popular open-source C language projects, QEMU and FFmpeg. 
This dataset consists of diverse function fragments. Then, Lu et al. \cite{lu2021codexglue} merges these projects and split them into a 8:1:1 training set, validation set, and test set.

\textbf{Reveal.} 
Reveal, constructed by Chakraborty et al. \cite{chakraborty2021deep}, extracts data from two large real-world C language open-source projects: Linux Debian Kernel and Chromium. 
The dataset has an imbalanced label distribution, with the number of non-vulnerable code snippets significantly exceeding the number of vulnerable code snippets. 
We also evenly divide Reveal into a training set, a validation set, and a test set in an 8:1:1 ratio.

\begin{table}[h!]
	\centering
	\caption{Overview of the dataset.}
	\resizebox{\linewidth}{!}{
		\begin{tabular}{c|c|c}
			\toprule[1pt]
			\textbf{Dataset}     & \textbf{Devign} & \textbf{Reveal} \\ 
			\midrule
			Description & \begin{tabular}[c]{@{}c@{}}Diverse vulnerability data constructed \\from FFmpeg and Qemu.\end{tabular}        & \begin{tabular}[c]{@{}c@{}}Real-world dataset constructed from\\ Linux Debian Kernel and Chromium.\end{tabular}       \\  
			\midrule
			Train       & 21854  & 18187  \\
			Valid       & 2732   & 2773   \\
			Test        & 2732   & 2774   \\
			Total       & 27318  & 22734  \\
			\midrule\midrule
			Vul         & 12460  & 2240   \\
			Non-Vul     & 14858  & 20494  \\ 
			\bottomrule[1pt]
		\end{tabular}
	}
	\label{dataset}
\end{table}

%\begin{figure*}[ht!]
%	\centering
%	\includegraphics[width=0.9\textwidth]{Fig4.pdf}
%	\caption{Vulnerability detection performance of the original model and the DFEPT-enhanced model on different data sets.}
%	\label{Fig4}
%\end{figure*}	
\vspace{-0.4cm}
\subsection{Baseline Methods}\label{4.2}
In our evaluation, we compare DFEPT with seven state-of-the-art methods.
\textbf{CodeBERT}\cite{feng2020codebert} 
%CodeBERT is a bimodal pre-trained model tailored for both natural language and programming language. 
%It can generate universal representations suitable for various natural language and programming language tasks, such as code understanding, clone detection, and so on.
stands as a bimodal pre-trained model designed for the nuanced processing of both natural and programming languages. It excels in crafting versatile representations that are well-suited for a broad spectrum of tasks across natural language and programming domains, including code comprehension, clone identification, among others.
\textbf{GraphCodeBERT}\cite{guo2020graphcodebert} 
extends the work of CodeBERT \cite{feng2020codebert} by incorporating code's data flow information into the Transformer architecture through encoding techniques. 
It is pre-trained with this integration, achieving improvements over the baselines set by CodeBERT.
\textbf{CodeT5}\cite{raffel2020exploring} 
%CodeT5 is an encoder-decoder model based on the T5 \cite{raffel2020exploring} architecture. 
%It introduces a new identifier-aware pre-trained task that enables the model to differentiate and recover masked identifiers. 
is built upon the T5 \cite{raffel2020exploring} architecture, functioning as an encoder-decoder model. 
It introduces an innovative identifier-aware pre-trained task, empowering the model to distinguish and restore obscured identifiers, thereby enhancing its understanding and processing of code.
Additionally, it proposes a bimodal code generation task, achieving better alignment between natural language and programming language.
\textbf{UniXcoder}\cite{guo2022unixcoder} 
expands upon the functionalities of models like CodeBERT \cite{feng2020codebert}, integrating a comprehensive understanding of code's syntax and semantics. 
This enhancement boosts the model's performance in downstream tasks such as code summarization, translation, and completion.
\textbf{ReGVD}\cite{nguyen2022regvd} 
proposes two novel techniques for constructing source code as graph structures, utilizing the embedding layer of GraphCodeBERT \cite{guo2020graphcodebert} for graph classification.
\textbf{VulBERTa}\cite{hanif2022vulberta} 
is a specialized pre-trained model for vulnerability detection, based on the RoBERTa \cite{liu2019roberta} architecture. It is suitable for multi-class vulnerability detection tasks.
\textbf{CSGVD}\cite{tang2023csgvd} 
proposes the use of pre-trained models and BiLSTM \cite{graves2005framewise} for dual embedding of Control Flow Graphs.
Additionally, it introduces a dual affine pooling layer and employs GCNs to classify source code.

\vspace{-0.2cm}
\subsection{Performance Metrics}\label{4.3}
We apply four widely recognized metrics in the software testing and analysis field \cite{zhou2019devign} to evaluate DFEPT.

\textbf{Accuracy.}
This metric represents the proportion of correctly classified samples to the total samples.
It can be calculated as: $Accuracy = \frac{TP+TN}{TP+TN+FP+FN}$, 
where TP, TN, FP, FN represent the number of true positives, true negatives, false positives, false negatives.

\textbf{Recall.}
This metric represents the proportion of correctly classified positive samples to actual positive samples.
It can be calculated as: $Recall = \frac{TP}{TP+FN}$

\textbf{Precision.}
This indicator quantifies the proportion of samples that are predicted to be positive among the samples that are actually positive.
It can be calculated as: $ Precision = \frac{TP}{TP+FP} $

\textbf{F1-Score.}
F1-Score is the harmonic mean of recall and precision, which reveals the robustness of the model.
It can be calculated as: $ F1 = 2 \times \frac{Precision \times Recall}{Precision+Recall} $

\vspace{-0.2cm}
\subsection{Research Questions}\label{4.4}
We establish the following three research questions (RQs) for evaluation. 

\begin{table*}[!ht]
	\centering
	\caption{Comparison of vulnerability detection results after combining different pre-trained models with DFEPT.}
	\resizebox{0.65\linewidth}{!}{
		\begin{tabular}{c||cccc|cccc}
			\toprule[1pt]
			\textbf{Dataset}             & \multicolumn{4}{c|}{\textbf{Devign}}              & \multicolumn{4}{c}{\textbf{Reveal}}              \\
			\midrule
			\textbf{Model}               & A\textbf{ccurancy} &\textbf{ F1}     & \textbf{Precision} & \textbf{Recall} & \textbf{Accuracy} & \textbf{F1}     & \textbf{Precision} & \textbf{Recall} \\
			\midrule\midrule
			CodeBERT            & 0.6332    & 0.471  & 0.698     & 0.3554 & 0.9085    & 0.3659 & 0.6122    & 0.2609 \\
			GraphCodeBERT       & 0.6384    & 0.4822 & \cellcolor{lb}\textbf{0.7044}    & 0.3665 & 0.9156    & 0.4074 & 0.7021    & 0.287  \\
			UniXcoder           & 0.6369    & 0.5466 & 0.6409    & 0.4765 & 0.9094    & 0.408  & 0.6017    & 0.3087 \\
			CodeT5              & 0.6336    & 0.5988 & 0.6024    & 0.5952 & 0.9265    & 0.4698 & 0.7048    & \cellcolor{lb}\textbf{0.3524} \\
			\midrule\midrule
			CodeBERT+DFEPT      & 0.6475    & 0.5523 & 0.6162    & 0.5004 & 0.9125    & 0.3801 & 0.6703    & 0.2652 \\
			GraphCodeBERT+DFEPT & 0.6296    & 0.5645 & 0.6137    & 0.5227 & 0.9265    & 0.4261 & \cellcolor{lb}\textbf{0.7654}    & 0.2952 \\
			UniXcoder+DFEPT     & \cellcolor{lb}\textbf{0.6497}    & 0.5726 & 0.6514    & 0.5108 & 0.9182    & 0.4464 & 0.7075    & 0.3261 \\
			CodeT5+DFEPT        & 0.6453    & \cellcolor{lb}\textbf{0.6122} & 0.615     & \cellcolor{lb}\textbf{0.6096} & \cellcolor{lb}\textbf{0.9292}    & \cellcolor{lb}\textbf{0.479}  & 0.7475    & \cellcolor{lb}\textbf{0.3524} \\
			\bottomrule[1pt]
		\end{tabular}
	}
	\label{RQ1}
\end{table*}

\textbf{RQ1: Can DFEPT improve the performance of pre-trained models?}
	
	In RQ1, we verify whether DFEPT can be combined with most code pre-trained models to enhance the performance of vulnerability detection.
	We integrate DFEPT with four popular code pre-trained models, using their embedding layers for node embedding and GCN for data flow embedding.
	We fine-tune both the original pre-trained models and the models combined with DFEPT on two datasets mentioned in Section \ref{4.1} . 
	These models are then applied to vulnerability detection tasks. 
	We report the performance metrics and the proportion of performance improvement for each model.

\textbf{RQ2: Can DFEPT outperform existing vulnerability detection methods?}
	
	In RQ2, we verify whether DFEPT can effectively detect vulnerabilities. 
%	We use the open-source code parsing tool tree-sitter \cite{tree-sitter} to parse the source code into AST representations, and then extract the data flow from the AST to embed it using a GCN. 
	For parsing source code into AST representations, we employ the open-source tool tree-sitter \cite{tree-sitter}. 
	Then, we extract the data flow from the AST and employ a GCN for embedding, thereby facilitating a deeper analysis and understanding of the code's structure and semantics.
	We report the best performance achieved by DFEPT and compare it with seven other baseline methods.

\textbf{RQ3: How do different graph embedding or pooling methods affect model performance?}

	In RQ3, we investigate the impact of different data flow embedding methods on the performance of DFEPT.
	For this study, we use UniXcoder as the base model. 
	During the node feature aggregation stage, we employ both GCN and GGNN for graph embedding. 
	In the pooling stage, we apply four pooling methods: max pooling, average pooling, sum pooling, and united pooling \cite{nguyen2022regvd}, each in combination with GCN and GGNN. 
	We compare the effects of different embedding methods on vulnerability detection and report the performance metrics.

In addition, we conduct \textbf{ablation studies} on DFEPT to understand the impact of each component on the overall performance.
We set up three sets of ablation experiments for DFEPT using UniXcoder as the base model to verify the necessity of each module. 
We remove the pre-trained model, data flow embedding, and sinusoidal positional encoding separately and observe the changes in the performance of the final model.

\vspace{-0.2cm}
\section{Experiment Results}
This section provides the experimental results and analyzes the performance of DFEPT, answering the research questions set in Section \ref{4.4}.

\vspace{-0.2cm}
\subsection{RQ1:Can DFEPT improve the performance of pre-trained models?}

Table \ref{RQ1} lists the accuracy, recall, precision and F1-Score of DFEPT combined with four pre-trained models.
Table \ref{RQ1-2} lists the relative improvement obtained by DFEPT compared with the only fine-tuned model.
%Further, Figure \ref{Fig4} provides a detailed comparison between the original model and the DFEPT-enhanced model.
Compared with most baselines, DFEPT achieves higher accuracy and F1-Score on all datasets. 
This shows that DFEPT can provide effective code structure information and data flow information to pre-trained models, thereby helping them better detect vulnerabilities.

\begin{table}[h!]
	\centering
	\caption{Vulnerability detection accuracy and F1-Score improvement percentage after different pre-trained models are combined with DFEPT.}
	\resizebox{\linewidth}{!}{
		\begin{tabular}{c||cc||cc}
			\toprule[1pt]
			\textbf{Dataset}                                                                & \multicolumn{2}{c||}{\textbf{Devign}}                                                                                                                  & \multicolumn{2}{c}{\textbf{Reveal}}                                                                                                                  \\
			\midrule
			\textbf{Models compared}                                                        & \textbf{\begin{tabular}[c]{@{}c@{}}Acc imp.(\%)\end{tabular}} & \textbf{\begin{tabular}[c]{@{}c@{}}F1 imp.(\%)\end{tabular}} & \textbf{\begin{tabular}[c]{@{}c@{}}Acc imp.(\%)\end{tabular}} & \textbf{\begin{tabular}[c]{@{}c@{}}F1 imp.(\%)\end{tabular}} \\
			\midrule\midrule
			\begin{tabular}[c]{c}CodeBERT vs.\\ CodeBERT+DFEPT\end{tabular}           & {2.26}                                                                & {17.26}                                                                 & {0.44}                                                                         & 3.88                                                                  \\
			\midrule
			\begin{tabular}[c]{c}GraphCodeBERT vs.\\ GraphCodeBERT+DFEPT\end{tabular} &  -1.38                                                                        &  17.07                                                                 &  1.19                                                                         &  4.59                                                                  \\
			\midrule
			\begin{tabular}[c]{c}UniXcoder vs.\\ UniXcoder+DFEPT\end{tabular}         &  2.01                                                                         &  4.76                                                                  &  0.97                                                                         &  9.41                                                                  \\
			\midrule
			\begin{tabular}[c]{c}CodeT5 vs.\\ CodeT5+DFEPT\end{tabular}               &  1.85                                                                         &  2.24                                                                  & 0.29                                                                         & 1.96       \\
			\bottomrule[1pt]                                                          
		\end{tabular}
	}
	\label{RQ1-2}
\end{table}
\begin{table*}[!ht]
	\centering
	\caption{Comparison of performance metrics DFEPT and other baselines.}
	\resizebox{0.65\linewidth}{!}{
		\begin{tabular}{c||cccc|cccc}
			\toprule[1pt]
			\textbf{Dataset}             & \multicolumn{4}{c|}{\textbf{Devign}}              & \multicolumn{4}{c}{\textbf{Reveal}}              \\
			\midrule
			\textbf{Model}               & \textbf{Accuracy} &\textbf{ F1}     & \textbf{Precision} & \textbf{Recall} & \textbf{Accuracy} & \textbf{F1}     & \textbf{Precision} & \textbf{Recall} \\
			\midrule\midrule
			CodeBERT            & 0.6332    & 0.471  & 0.698     & 0.3554 & 0.9085    & 0.3659 & 0.6122    & 0.2609 \\
			GraphCodeBERT       & 0.6384    & 0.4822 & \cellcolor{lb}\textbf{0.7044}    & 0.3665 & 0.9156    & 0.4074 & 0.7021    & 0.287  \\
			UniXcoder           & 0.6369    & 0.5466 & 0.6409    & 0.4765 & 0.9094    & 0.408  & 0.6017    & 0.3087 \\
			CodeT5              & 0.6336    & 0.5988 & 0.6024    & 0.5952 & 0.9265    & 0.4698 & 0.7048    & 0.3524 \\
			ReGVD           & 0.6285    & 0.5955 & 0.5957    & 0.5952 & 0.9015    & 0.3946  & 0.5214    & 0.3174 \\
			VulBERTa        & 0.6413    & 0.6045 & 0.6124    & 0.5968 & 0.8646    & 0.4597  & 0.3853    & \cellcolor{lb}\textbf{0.5696} \\
			CSGVD           & 0.6446    & 0.6039 & 0.6187    & 0.5899 & 0.9008    & 0.3297  & 0.5114    & 0.2432 \\
			\midrule
			DFEPT        & \cellcolor{lb}\textbf{0.6453}    & \cellcolor{lb}\textbf{0.6122} & 0.615     & \cellcolor{lb}\textbf{0.6096} & \cellcolor{lb}\textbf{0.9292}    & \cellcolor{lb}\textbf{0.479}  & \cellcolor{lb}\textbf{0.7475}    & 0.3524 \\
			\bottomrule[1pt]
		\end{tabular}
	}
	\label{RQ2}
\end{table*}

When DFEPT is combined with CodeT5, we obtain the best F1 scores on both datasets, which shows that the embedding layers initialized by different pre-trained models have a greater impact on the data flow embedding of DFEPT. 
Concurrently, it has been observed that DFEPT offers limited enhancements to more advanced baseline methods. 
This may be attributed to the fact that superior baseline models employ more sophisticated pre-training techniques and possess larger parameter sizes, resulting in a diminished proportion of effective data flow information that DFEPT can provide.
It is worth noting that when DFEPT is combined with GraphCodeBERT, the accuracy on Devign decreases. 
This may be because GraphCodeBERT has already used data flow information once during pre-trained, and repeated data flow information aggravates the overfitting of the model.
DFEPT can improve the F1-Score of all pre-trained models to a certain extent, which proves that the model combined with DFEPT has stronger robustness.

%\begin{tcolorbox}[top=5pt, bottom=5pt,colback=gray!20!white, colframe=black,boxrule=1pt]
%	\textbf{Answer to RQ1:} 
%	DFEPT can be combined with most pre-trained models to steadily improve the accuracy and robustness of vulnerability detection. 
%	The degree of improvement is affected by the pre-trained model itself and the embedding layer.
%\end{tcolorbox}

\aptLtoX[graphic=no,type=html]{
\begin{shaded}
\noindent
%\fcolorbox{black}{gray!20}{
	\parbox{\dimexpr\linewidth-3\fboxsep-2\fboxrule\relax}{%
		\vspace{5pt}% Top padding
		\textbf{Answer to RQ1:}
		
	DFEPT can be combined with most pre-trained models to steadily improve the accuracy and robustness of vulnerability detection. 
	The degree of improvement is affected by the pre-trained model itself and the embedding layer.
		\vspace{5pt}% Bottom padding
	}
\end{shaded}
}{
\noindent
\fcolorbox{black}{gray!20}{
	\parbox{\dimexpr\linewidth-3\fboxsep-2\fboxrule\relax}{%
		\vspace{5pt}% Top padding
		\textbf{Answer to RQ1:}
	DFEPT can be combined with most pre-trained models to steadily improve the accuracy and robustness of vulnerability detection. 
	The degree of improvement is affected by the pre-trained model itself and the embedding layer.
		\vspace{5pt}% Bottom padding
	}
}}

\subsection{RQ2: Can DFEPT outperform existing vulnerability detection methods?}

As is shown in Table \ref{RQ2}, we compare the results of CodeT5+DFEPT with the remaining seven baseline methods on two datasets.
The results in Table \ref{RQ2} show that on the Devign dataset, DFEPT leads all baseline methods with an accuracy of 64.53\% and an F1-Score of 61.22\%. 
This shows that DFEPT has the most balanced performance and strongest robustness in correctly identifying vulnerabilities, and does not tend to over-predict and over-conservative prediction situations. 
On the Reveal dataset, DFEPT again exceeds all baselines with an accuracy of 92.92\% and an F1-Score of 47.9\%. This shows that DFEPT has the excellent generalization ability.

The performance improvement of DFEPT on Reveal is sometimes less than that of Devign. 
We believe this is due to the imbalance of the Reveal dataset. 
Due to the limited quantity and diversity of vulnerable codes in Reveal, many data flow information containing vulnerabilities have not been fully learned.
As a result, it cannot provide sufficient semantic information for vulnerability detection.
%Although DFEPT only uses data types as node features for embedding, this data flow embedding does provide effective structural information for vulnerability detection. 

Compared to sequence-based models, sequence-based models generally have poorer predictive performance due to their inability to effectively perceive and utilize the structured information of code, such as control flows and data flows, resulting in insufficient vulnerability features. 
In contrast, directly using Graph Neural Networks to aggregate node features can lead to the loss of the model's ability to perceive the contextual environment. 
Therefore, the practice of using data flow embedding to provide additional effective features for sequence-based models results in higher and more balanced overall performance in vulnerability detection.

%\begin{tcolorbox}[top=5pt, bottom=5pt,colback=gray!20!white, colframe=black,boxrule=1pt]
%	\textbf{Answer to RQ2:} 
%	DFEPT outperforms all baseline methods. It effectively provides semantic information to pre-trained models, achieving higher performance in vulnerability detection tasks under various experimental conditions.
%\end{tcolorbox}
\aptLtoX[graphic=no,type=html]{\begin{shaded}
\noindent
	\parbox{\dimexpr\linewidth-3\fboxsep-2\fboxrule\relax}{%
		\vspace{5pt}% Top padding
		\textbf{Answer to RQ2:}
	DFEPT outperforms all baseline methods. It effectively provides semantic information to pre-trained models, achieving higher performance in vulnerability detection tasks under various experimental conditions.
		\vspace{5pt}% Bottom padding
	}
\end{shaded}}{
\noindent
\fcolorbox{black}{gray!20}{
	\parbox{\dimexpr\linewidth-3\fboxsep-2\fboxrule\relax}{%
		\vspace{5pt}% Top padding
		\textbf{Answer to RQ2:}
	DFEPT outperforms all baseline methods. It effectively provides semantic information to pre-trained models, achieving higher performance in vulnerability detection tasks under various experimental conditions.
		\vspace{5pt}% Bottom padding
	}
}}
\vspace{-0.2cm}
\subsection{RQ3: How do different graph embedding or pooling methods affect model performance?}

\begin{table*}[!ht]
	\centering
	\caption{Performance of DFEPT under different settings.}
	\resizebox{0.65\linewidth}{!}{
		\begin{tabular}{c||cccc|cccc}
			\toprule[1pt]
			\multirow{2}{*}{\textbf{Dataset} }& \multicolumn{4}{c|}{\textbf{Devign}}                                     & \multicolumn{4}{c}{\textbf{Reveal}}                                     \\
			\cmidrule{2-9}
			& \textbf{Accuracy} & \textbf{F1} & \textbf{Precision} & \textbf{Recall} & \textbf{Accuracy} & \textbf{F1} & \textbf{Precision} & \textbf{Recall} \\
			\midrule\midrule
			GGNN-sum         & 0.6274             & 0.6005      & 0.5916             & 0.6096          & 0.9099             & 0.3653      & 0.6344             & 0.2565          \\
			GCN-sum          & 0.6402             & 0.6165      & 0.604              & 0.6295          & 0.9195             & 0.4404      & 0.7423             & 0.313           \\
			\midrule
			GGNN-max         & 0.6391             & 0.6216      & 0.5996             & 0.6454          & 0.9116             & 0.3927      & 0.6436             & 0.2826          \\
			GCN-max          & 0.638              & 0.6171      & 0.6002             & 0.6351          & 0.9103             & 0.4205      & 0.6066             & 0.3217          \\
			\midrule
			GGNN-mean        & 0.6369             & 0.619       & 0.5975             & 0.6422          & 0.9116             & 0.3186      & 0.7231             & 0.2043          \\
			GCN-mean         & 0.6354             & 0.5193      & 0.6585             & 0.4287          & 0.9103             & 0.3818      & 0.63               & 0.2739          \\
			\midrule
			GGNN-uni         & 0.6318             & 0.6221      & 0.5885             & 0.6598          & 0.9085             & 0.3918      & 0.5982             & 0.2913          \\
			GCN-uni          & 0.6497             & 0.5726      & 0.6514             & 0.5108          & 0.9182             & 0.4464      & 0.7075             & 0.3261         \\
			\bottomrule[1pt]
		\end{tabular}
	}
	\label{RQ3}
\end{table*}

As shown in Table \ref{RQ3}, we use UniXcoder \cite{guo2022unixcoder} as the base model. 
The choice of UniXcoder is due to its ability to efficiently fine-tune with fewer samples, and its encoder mode, which facilitates better integration of data flow embedding vectors \cite{guo2022unixcoder}. 
We apply both GCN and GGNN for graph embedding and use four common pooling methods, resulting in a total of eight combinations. 
The results indicate that regardless of the combination of data flow embedding, each can effectively embed data flow to enhance the performance of the vulnerability detection model.

\begin{figure}[h!]
	\centering
	\resizebox{0.45\textwidth}{!}{\includegraphics[width=1\textwidth]{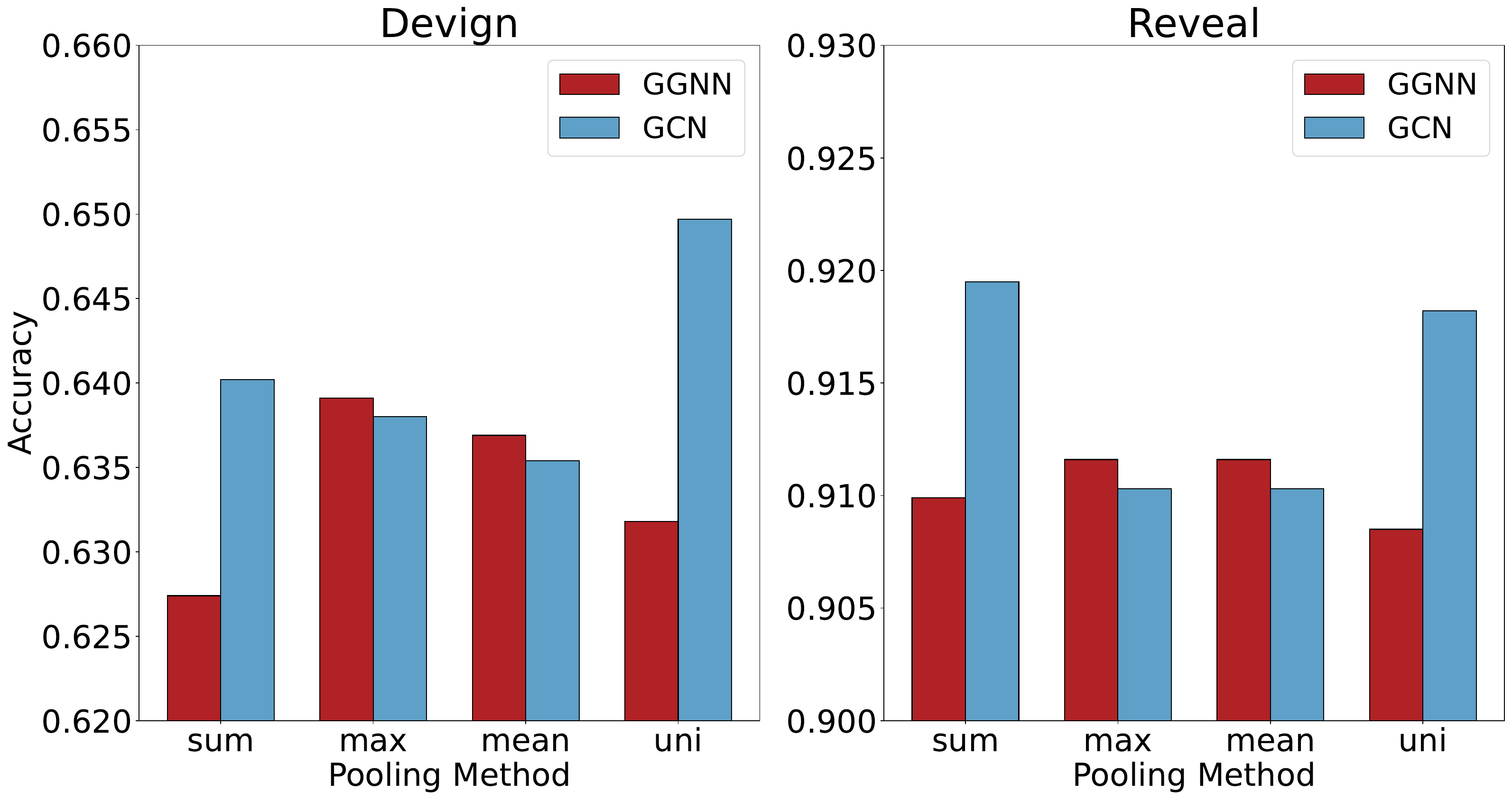}}
	\caption{Accuracy with different settings of DFEPT.}
	\label{Fig3}
\end{figure}

We plot bar graphs of the model accuracy achieved with different graph embedding and pooling methods on the Devign and Reveal datasets, respectively. 
As shown in Figure \ref{Fig3}, the fluctuation in accuracy exhibits similar trends on both datasets. 
Under max pooling and average pooling, the performance achieved with GCN embedding and GGNN embedding of the data flow is not significantly different. 
United pooling combined with GCN yields the best performance on Devign. 
Under sum pooling and united pooling, the performance achieved with GCN embedding is higher than with GGNN embedding. 
This is because united pooling combines the advantages of sum pooling, which allows GNN to access all available information, leading to a more effective embedding result \cite{nguyen2022regvd,xu2018powerful}.
Additionally, GGNN aggregates information from neighboring nodes through a gating mechanism, which is effective in capturing long-term dependencies. 
However, since vulnerabilities are typically focused on a specific data flow or a local data feature, GGNN's ability to embed data flows is not as effective as GCN's.
%\begin{tcolorbox}[top=5pt, bottom=5pt,colback=gray!20!white, colframe=black,boxrule=1pt]
%	\textbf{Answer to RQ3:} 
%	
%	Regardless of the combination used for data flow embedding, it can provide effective data flow information for the pre-trained model. 
%	When DFEPT uses GCN with united pooling for data flow embedding, the vulnerability detection model achieves the state-of-the-art performance.
%\end{tcolorbox}

\aptLtoX[graphic=no,type=html]{\begin{shaded}
\noindent
%\fcolorbox{black}{gray!20}{
	\parbox{\dimexpr\linewidth-3\fboxsep-2\fboxrule\relax}{%
		\vspace{5pt}% Top padding
		\textbf{Answer to RQ3:}
	Regardless of the combination used for data flow embedding, it can provide effective data flow information for the pre-trained model. 
	When DFEPT uses GCN with united pooling for data flow embedding, the vulnerability detection model achieves the state-of-the-art performance.
		\vspace{5pt}% Bottom padding
	}
\end{shaded}}{
\noindent
\fcolorbox{black}{gray!20}{
	\parbox{\dimexpr\linewidth-3\fboxsep-2\fboxrule\relax}{%
		\vspace{5pt}% Top padding
		\textbf{Answer to RQ3:}
	Regardless of the combination used for data flow embedding, it can provide effective data flow information for the pre-trained model. 
	When DFEPT uses GCN with united pooling for data flow embedding, the vulnerability detection model achieves the state-of-the-art performance.
		\vspace{5pt}% Bottom padding
	}
}}

\subsection{Ablation Study}
In order to comprehensively understand the impact of each module in our proposed model, we conduct an ablation experiment using UniXcoder as the base model.
The ablation experiment aims to study the contribution of the three components of pre-trained model, GNN, and sinusoidal position encoding to the overall performance of the vulnerability detection model. 
We test the performance of the model on two datasets, Devign and Reveal, and the results are shown in Table \ref{ablation}.

\begin{table}[!h]
	\centering
	\caption{Ablation study evaluated on the test dataset.}
	\resizebox{1\linewidth}{!}{
		\begin{tabular}{cc||cccc}
			\toprule
			\multicolumn{2}{c||}{\textbf{Model}}                 & \textbf{Accuracy}                                & \textbf{F1}                                                             & \textbf{Precision}                               & \textbf{Recall}                                  \\
			\midrule
			& DFEPT w/o GNN             & 0.6369                                  & 0.5466                                                         & 0.6409                                  & 0.4765                                  \\
			& DFEPT w/o pre-trained Model                   & 0.5458                                  & 0.2929                                                         & 0.514                                   & 0.2048                                  \\
			& DFEPT w/o Sin. Encoding & 0.6417                                  & 0.5305                                                         & \cellcolor{lb}{\textbf{0.6663}} & 0.4406                                  \\
			\multirow{-4}{*}{\rotatebox{90}{Devign}}& DFEPT          & \cellcolor{lb}{\textbf{0.6497}}   & \cellcolor{lb}{\textbf{0.5726}} & 0.6514                                & \cellcolor{lb}{\textbf{0.5108}} \\
			\midrule\midrule
			& DFEPT w/o GNN            & 0.9094                                  & 0.408                                                          & 0.6017                                  & 0.3087                                  \\
			& DFEPT w/o pre-trained Model                  & 0.8989                                  & 0.0417                                                         & 0.5                                     & 0.0217                                  \\
			& DFEPT w/o Sin. Encoding & 0.9164                                  & 0.4412                                                         & 0.6818     &  0.3261    \\ 
			\multirow{-4}{*}{\rotatebox{90}{Reveal}}& DFEPT      & \cellcolor{lb}{\textbf{0.9182}}   & \cellcolor{lb}{\textbf{0.4464}}& \cellcolor{lb}{\textbf{0.7075}}    & \cellcolor{lb}{\textbf{0.3261}} \\
			\bottomrule
		\end{tabular}
	}
	\label{ablation}
\end{table}

\textbf{GNN.}
GNN is an indispensable component for processing data flow graphs. 
Without GNN, data flow information cannot be input into the model, causing it to revert to a mere pre-trained model. 
On the Devign dataset, the accuracy and F1-Score of the model decrease by 1.28\% and 2.6\%, respectively.
Similar declines are observed on Reveal, with reductions of 0.88\% and 3.84\%. 
This indicates that GNN is effective and necessary for aggregating graph features.

\textbf{Pre-trained model.}
DFEPT relies on the pre-trained model for vulnerability detection. 
Removing the pre-trained model results in a decrease of 10.39\% in accuracy and 27.97\% in F1-Score on Devign, and a decline of 1.93\% and 40.47\% on Reveal. 
This drastic drop highlights the excellent ability of the pre-trained model in sequence processing and also points out the limitations of DFEPT as a supplementary model to the pre-trained model, namely that it cannot independently complete the task of vulnerability detection.

\textbf{Sinusoidal position encoding.}
Sinusoidal position encoding is intended to provide the position information of the data flow itself to the vulnerability detection model. 
When this positional information is removed, the model's performance only suffers a minor decline. 
On Devign, the accuracy and F1-Score decrease by 0.8\% and 4.21\%, respectively, and on Reveal, the decrease is 0.18\% and 0.52\%. 
This suggests that while sinusoidal position encoding can provide effective vulnerability information, it is not as indispensable as GNN.

\section{Discussion}
In this section, we will discuss the reason why DFEPT work and threats to validity.

\vspace{-0.3cm}
\subsection{Why does DFEPT  Work?}
DFEPT achieves remarkable performance in aggregating data flow information without using segmented statement-level text or token information, such as assignment statements. 
This success is likely because, in the data flow graph structure, the key to generating vulnerabilities lies in determining the direction of data flow and the process of variable transition, rather than variable naming or assignment statements. 
Therefore, using simple data types to identify different data flows within a code segment effectively accomplishes semantic and structural embedding crucial for vulnerability detection.

When combined with pre-trained models, DFEPT achieves the highest performance, indicating that pre-trained models relying solely on textual information may not sufficiently learn data flows. 
Thus, DFEPT supplements the necessary information for vulnerability detection. 
DFEPT integrates well with most pre-trained models because it does not depend on potential text-level spurious features, such as variable and function names. 
These features are not directly related to the generation of vulnerabilities, so including them as a part of the input might lead to model errors.

\begin{figure}[h!]
	\centering
	\resizebox{0.4\textwidth}{!}{\includegraphics[width=1\textwidth]{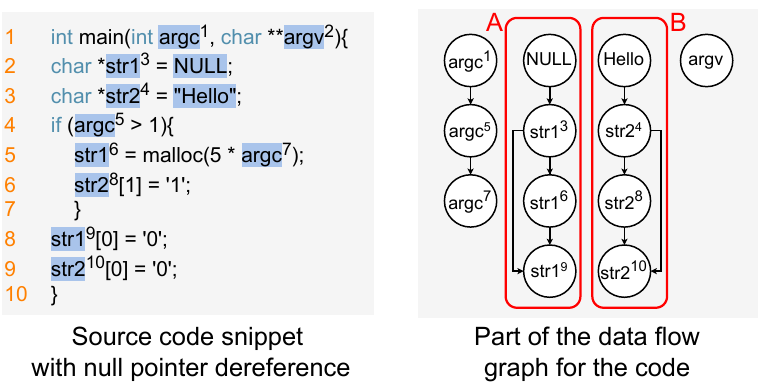}}
	\caption{An example explaining the effectiveness of how DFEPT works.}
	\label{Fig5}
\end{figure}

DFEPT uses sinusoidal positional encoding, proposed in Transformer structures for extracting sequence position, to embed data flow structures. 
This approach is not only to match the sequence-based training mode of pre-trained models but also to input features like the positional and directional information of data flows. 
This is because the same data flow in different positions of a code segment could lead to different outcomes.
For example, figure \ref{Fig5} presents a specific code fragment that exhibits a vulnerability due to dereferencing of a null pointer. 
Specifically, within the data flow A highlighted in the figure, a pathway exists wherein a null pointer, not assigned any memory, is directly utilized ($NULL \to str1^3 \to str1^9$). 
Through graph learning, the semantic representation of this vulnerability is embedded into the graph vector. 
Further, the data flow B, also annotated in figure \ref{Fig5}, bears a striking resemblance to data flow A but does not contain the vulnerability, attributed to the initial assignment of $str2$ not being $NULL$. 
Sinusoidal positional encoding furnishes the resultant graph vector with relative positional information, thereby distinguishing between data flows A and B. 
The data flow semantic information provided by DFEPT enables the pre-trained model to effectively identify vulnerabilities propagated through data flow. 
Thus, leveraging the data flow semantic information furnished by DFEPT, the pre-trained model can efficaciously recognize vulnerabilities resulting from data flow propagation.

\vspace{-0.3cm}

\subsection{Threats to Validity}
\subsubsection{Internal Validity.}
The main threats to internal validity come from the choice of pre-trained models and different code graph structures.
The performance of pre-trained models affects DFEPT's performance because DFEPT uses the embedding layer of pre-trained models when embedding and encoding data flows. 
Different pre-trained models have distinct pre-trained tasks, and their embedding layers might be more biased towards generative tasks rather than classification. 

DFEPT only uses a single data flow graph as input, which may limit the performance of DFEPT. 
Abstract syntax trees, control flow graphs and program dependency graphs also play an extremely important role in code analysis and may contain structural information that data flow graphs do not have.
Moreover, one possible reason for DFEPT's inefficacy is that data flow analysis cannot identify all types of vulnerabilities, as some are not generated by code data flows, such as insufficient authentication (CWE-287).
\vspace{-0.2cm}
\subsubsection{External Validity.}
We primarily evaluate DFEPT on the Devign \cite{zhou2019devign} and Reveal \cite{chakraborty2021deep} datasets because all baseline models support these datasets. 
%Both datasets are extracted from open-source projects written in C language. 
%When detecting vulnerabilities in projects written in other programming languages, additional noise might be introduced due to differences in syntax and keywords. 
These datasets are derived from open-source projects coded in the C language. It's important to note that when extending vulnerability detection to projects developed in other programming languages, potential discrepancies in syntax and keywords could introduce extraneous noise, affecting the detection process.
Additionally, in real-world vulnerability detection scenarios, we often face extremely imbalanced data, characterized by a small proportion of vulnerable samples and a diversity of vulnerability types. 
This imbalance may prevent DFEPT from extracting sufficiently representative data flow features, thus failing to obtain effective structural information.
\vspace{-0.3cm} 
\section{Related Work}
In this section, we will introduce the existing approaches for software vulnerability detection, including traditional detection techniques, deep learning-based approaches, and pre-trained model-based approaches.
\vspace{-0.3cm}

\subsection{Traditional Vulnerability Detection}
Traditional software vulnerability detection primarily relies on domain experts manually creating extensive rule libraries \cite{dfh25, checkmarx} for detection. 
By applying semi-automatic techniques such as taint tracking \cite{kang2022tracer}, symbolic execution \cite{wang2020wana}, fuzzing \cite{dinh2021favocado}, and code similarity-based matching \cite{sun2021vdsimilar}, the efficiency of detection is enhanced. 
Although theses approaches achieve vulnerability detection, they suffer from a lack of complete automation and high rule formulation costs. 
Additionally, it may lead to increased rates of false positives and false negatives \cite{yamaguchi2015pattern}.

\vspace{-0.3cm}
\subsection{Deep Learning-Based Vulnerability Detection}
The success of Deep Learning (DL) technologies in various fields has motivated researchers to apply DL techniques for automated vulnerability detection \cite{chakraborty2021deep}.
Existing DL-based vulnerability detection models primarily fall into two categories: sequence-based models and models based on code graph structures.

Sequence-based models treat code as natural language sequences for modeling, extracting vulnerability features from code snippets to achieve vulnerability detection.
For instance, Li et al. \cite{li2018vuldeepecker} define "code gadgets" from the perspective of code slicing, and they extract and assemble library/API function calls. They then apply a Bidirectional Long Short-Term Memory (BiLSTM) \cite{graves2005framewise} model for detection.

Graph-based methods model the source code using graph data structures, followed by the utilization of Graph Neural Networks (GNNs) for vulnerability detection.
Commonly used code graph structures include ASTs, CFGs, DFGs, and PDGs, among others.
Recently, Nguyen et al. \cite{nguyen2022regvd} propose two graph construction techniques: unique token-focused and index-focused construction.
They construct graphs from source code sequences and implement vulnerability detection using Residual Graph Convolutional Networks.
Tang et al. \cite{tang2023csgvd} perform graph node feature embedding of CFGs by using pre-trained model and BiLSTM.
These embedded features are then inputted into a graph convolutional layer for detection.

\vspace{-0.3cm}
\subsection{Pre-trained Model-Based Vulnerability Detection}
%In recent years, pre-trained language models have increasingly been integrated into a variety of application scenarios, with researchers exploring the use of these models to enhance the accuracy of vulnerability detection \cite{steenhoek2023empirical}.
In recent times, the integration of pre-trained language models into diverse application contexts has seen a notable uptick, with scholars investigating how these models can be leveraged to improve the precision in identifying vulnerabilities \cite{steenhoek2023empirical}.
The core concept of pre-trained models involves initially conducting pre-trained on a vast amount of source code data, followed by fine-tuning for specific downstream tasks \cite{kanade2020learning}.

Feng et al. \cite{feng2020codebert} propose CodeBERT, which combines the processing capabilities of natural language and programming language, specifically designed for understanding and generating source code. 
%CodeBERT has demonstrated strong performance in downstream tasks such as vulnerability detection, code search, and code comment generation.
%To address the limitation of CodeBERT's lack of code structure awareness, Guo et al. \cite{guo2020graphcodebert} propose a new pre-trained approach that encodes data flow and integrates it into the Transformer \cite{vaswani2017attention} architecture, leading to an improved performance. 
Wang et al. \cite{wang2021codet5} present CodeT5, an encoder-decoder model based on the T5 \cite{raffel2020exploring} architecture, which exhibits enhanced code structure perception capabilities. 
Hanif et al. \cite{hanif2022vulberta}, building on the RoBERTa \cite{liu2019roberta} architecture, conduct retraining to develop VulBERTa, a pre-trained model specifically for vulnerability detection, achieving precise detection of vulnerabilities.

%However, if pre-trained models are directly applied to vulnerability detection after fine-tuning, they face the challenge of capturing vulnerability features from lengthy codes and complex structures \cite{zhang2023vulnerability}. 
%Additionally, due to the diverse nature of different vulnerabilities, it is challenging to extract useful detection features from the code representation level.

Furthermore, researchers has explored some methods of providing code structure information to pre-trained models, such as methods based on program execution paths \cite{10153647} and distribution based on the number of vulnerabilities \cite{10497542}.

Zhang et al. \cite{10153647}  achieve CodeBERT's understanding of the code structure by providing CodeBERT with program execution path information on CFG.
Wen et al. \cite{10497542} observe the long-tail distribution characteristics of vulnerability types and implement adaptive learning of the difference between head and tail classes.
\vspace{-0.3cm}

\section{Conclusion And Future Work}
We propose DFEPT, an efficient data flow embedding technique designed to enhance vulnerability detection in models based on pre-trained models. 
Our data flow embedding starts from two core aspects: structural information and positional information, aiming to uncover potential vulnerability formation patterns and characteristics. 
We use data types to identify different data flows within the same code segment, where each data flow could potentially capture the cause of a vulnerability. 
DFEPT combines graph embedding and sinusoidal positional encoding with transformer architecture pre-trained models for vulnerability detection and is applicable to most pre-trained models. 
Our experimental results demonstrate that DFEPT is highly effective, improving the vulnerability detection performance of all pre-trained models, and outperforming all baseline methods. 
We also investigate the impact of different graph embedding methods on DFEPT, finding that all embedding methods enhance the vulnerability detection performance of pre-trained models.
In the future, we plan to extend DFEPT to ASTs, CFGs, and PDGs to explore graph embedding methods that can provide more effective information. 
We also intend to explore the application of interpretability tools to precisely locate the lines of code where vulnerabilities arise and assess the efficacy of our technique across different programming languages.
\vspace{-0.2cm}
\begin{acks} 
%We would like to thank anonymous reviewers for their insightful and constructive comments.
This work was supported in part by the National Natural Science Foundation of China (No. 62372071), the Chongqing Technology Innovation and Application Development Project (No. CSTB2022TIAD-STX0007 and No. CSTB2023TIAD-STX0025), the Fundamental Research Funds for the Central Universities (No. 2023CDJKYJH013), and the National Training Program of Innovation and Entrepreneurship for Undergraduates (No. 202310611109).
\end{acks}

\balance
\bibliographystyle{ACM-Reference-Format}
\bibliography{references}

\end{document}